\documentclass[showpacs,aps,graphicx,twocolumn]{revtex4}%
\usepackage{graphicx}
\usepackage{multirow}

\usepackage{amssymb}
\usepackage{amsmath}
\usepackage{amsfonts}

\begin{document}


\title{Error-detected generation and complete analysis of hyperentangled
Bell states for photons assisted by quantum-dot spins in
double-sided optical microcavities\footnote{Published in Opt.
Express \textbf{24}, 28444--28458  (2016)}}

\author{Guan-Yu Wang,$^{1}$ Qing Ai,$^{1}$ Bao-Cang Ren,$^{2}$ Tao Li,$^{1}$ and Fu-Guo Deng$^{1,}$\footnote{Corresponding author: fgdeng@bnu.edu.cn} }

\address{$^{1}$Department of Physics, Applied Optics Beijing Area Major Laboratory,
Beijing Normal University, Beijing 100875, China\\
$^{2}$Department of Physics, Capital Normal University, Beijing
100048, China}

\date{\today }

\begin{abstract}
We construct an error-detected block, assisted by the quantum-dot
spins in double-sided optical microcavities. With this block, we
propose three error-detected schemes for the deterministic
generation, the complete analysis, and the complete nondestructive
analysis of hyperentangled Bell states in both the polarization and
spatial-mode degrees of freedom of two-photon systems. In these
schemes, the errors can be detected, which can improve their
fidelities largely, far different from other previous schemes
assisted by the interaction between the photon and the
quantum-dot-cavity system. Our scheme for the deterministic
generation of hyperentangled two-photon systems  can be performed by
repeat until success. These features make our schemes more useful in
high-capacity quantum communication with hyperentanglement in the
future.
\end{abstract}

 \maketitle

\section{Introduction}

Quantum entanglement plays a critical role in quantum information
processing \cite{quantum1} and it is a key quantum resource in
quantum communication, such as quantum teleportation \cite{tele1},
quantum dense coding \cite{dense1,super2}, quantum key distribution
\cite{Ekert}, quantum secret sharing \cite{QSS1}, quantum secure
direct communication \cite{QSDC1,twostep}, and so on. The complete
and deterministic analysis of Bell states is required to achieve
some important tasks in quantum communication. In 1999, two schemes
of Bell-state analysis (BSA) \cite{BSA1,BSA2} for teleportation with
only linear optical elements were proposed. However, it is
impossible to deterministically and completely distinguish the four
Bell states in polarization with only linear optical elements
\cite{BSA3,BSA4,BSA5}. In 2005, Barrett \emph{et al.} \cite{BSA9}
proposed an analyzer for distinguishing all four polarization Bell
states using weak nonlinearities. In 2015, Sheng \emph{et al.}
\cite{LBSA} realized the near complete logic Bell-state analysis by
using the cross-Kerr nonlinearity. The complete BSA can be
accomplished with the assistance of hyperentanglement
\cite{BSA7,BSA8,hyper9,hyper8,DEPPTS}. In 1998, Kwiat and Weinfurter
\cite{BSA7} introduced the method to distinguish the four Bell
states of photon pairs in the polarization degree of freedom (DOF)
with the assistance of the hyperentanglement in both the
polarization DOF and the momentum DOF. In 2003, Walborn \emph{et
al.} \cite{BSA8} proposed a simple linear-optical scheme for the
complete Bell-state measurement of photons by using hyperentangled
states with linear optics. In 2006, Schuck \emph{et al.}
\cite{hyper9} deterministically distinguished polarization Bell
states of entangled photon pairs completely assisted by
polarization-time-bin hyperentangled states in experiment. In 2007,
Barbieri \emph{et al.} \cite{hyper8} realized a complete and
deterministic Bell-state measurement using linear optics and
two-photon hyperentangled in polarization and momentum DOFs in
experiment.

Besides the assistance for the complete BSA, hyperentanglement, a
state of a quantum system being simultaneously entangled in multiple
DOFs, has attracted much attention as it can improve the channel
capacity largely, beat the channel capacity of linear photonic
superdense coding \cite{super1}, and can be used for quantum
error-correcting code \cite{correct} and quantum repeaters
\cite{repeater2}. Some theoretical and experimental schemes for the
generation of hyperentangled states have been proposed and
implemented in optical systems
\cite{hyper1,hyper2,hyper3,hyper4,hyper5,hyper6}, such as
polarization-momentum DOFs \cite{hyper2},
polarization-orbital-angular momentum DOFs \cite{hyper4}, multipath
DOFs \cite{hyper5}, and so on. In 2009, Vallone \emph{et al.}
\cite{hyper6} demonstrated experimentally the generation of a
two-photon six-qubit hyperentangled state in three DOFs. A
hyperentangled photon pair, which is hyperentangled in both the
polarization and spatial-mode DOFs with $16$ orthogonal Bell states,
can be produced by spontaneous parametric down-conversion source
with the $\beta$  barium  borate  crystal. However the quantum
efficiency of this method is low and the the multiphoton generation
probability is high, which will limit the application of
hyperentanglement in quantum information processing. Moreover, one
cannot distinguish the $16$ Bell states completely with only linear
optics. In 2010, Sheng  \emph{et al.} \cite{HBSA1} proposed the
first scheme for the complete hyperentangled BSA (HBSA) of all the
two-photon polarization-spatial hyperentangled states with
cross-Kerr nonlinearity. In 2016, Li \emph{et al.} \cite{HBSA2}
presented a simplified complete HBSA with cross-Kerr nonlinearity.
As the solid state system can provide giant nonlinearity, it is
viewed as a promising candidate for quantum information processing
and quantum computing. In 2012, Ren \emph{et al.} \cite{HBSA3}
proposed a scheme for complete HBSA assisted by quantum-dot spins in
a one-side optical microcavity. In the same year, Wang \emph{et al.}
\cite{HBSA4} proposed two interesting schemes for
hyperentangled-Bell-state generation (HBSG) and HBSA by quantum-dot
spins in double-sided optical microcavities. In 2015, Liu \emph{et
al.} \cite{HBSA5} proposed two schemes for HBSG and HBSA assisted by
nitrogen-vacancy centers in resonators.

The electron spin in a GaAs-based or InAs-based charged quantum dot
(QD) is an attractive candidate for solid-state quantum information
processing. The electron-spin coherence time of a charged QD can be
maintained for more than $3\mu s$ \cite{QD1,QD2} and the electron
spin-relaxation time can be longer ($\sim ms$) \cite{QD3,QD4}.
Moreover, it is comparatively easy to embed the QDs in the
solid-state cavities and it can be easily manipulated fast and
initialized \cite{QD5,QD7,QD8}. Based on a singly charged QD inside
an optical resonant cavity, many interesting schemes for quantum
information processing have been proposed
\cite{HBSA3,HBSA4,QD9,QD10,QD12,QD13,QD14,QD15}. In the ideal
condition, the fidelities and the efficiencies of these schemes can
be $100\%$. In realistic condition, their fidelities and the
efficiencies  would be affected by the parameters of the QD-cavity
systems more or less. In 2011, Kastorynao \emph{et al.}
\cite{Sorensen} proposed a scheme for the preparation of a maximally
entangled state utilizing the decay of the cavity to improve the
fidelity, which can herald the error. In 2012, Li \emph{et al.}
\cite{Li} proposed a robust-fidelity scheme for atom-photon
entangling gates, which is adequate for high-fidelity maximally
entangling gates even in the weak-coupling regime based on the
scattering of photons off single emitters in one-dimensional
waveguides.

In this paper, we construct an error-detected block assisted by the
QD spins in double-sided optical microcavities. With this block, we
propose three schemes for the deterministic HBSG,  the complete
HBSA, and the complete nondestructive HBSA of hyperentangled Bell
states in both the polarization and spatial-mode DOFs of two-photon
systems. As the errors can be detected in these schemes, they
possess the advantage of having high fidelity, far different from
other previous schemes by others, especially in our scheme for the
deterministic generation of hyperentangled two-photon systems as it
can be performed by repeat until success. Moreover, our schemes work
in both the weak coupling regime and the strong coupling regime.
These features make our schemes more useful in high-capacity quantum
communication with hyperentanglement in the future.

\begin{figure*}[ht!]
\centering\includegraphics[width=12cm]{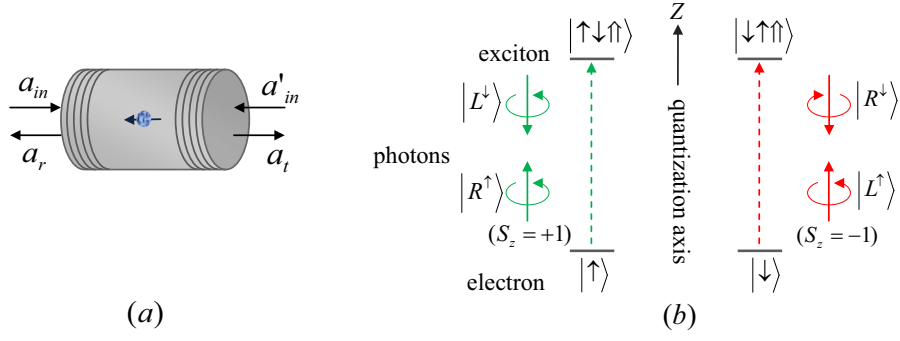} \caption{(a) A
schematic diagram for a singly charged QD inside a double-sided
optical microcavity. (b) Schematic description of the relevant
exciton energy levels and the spin selection rules for optical
transition of negatively charged excitons. $|R^{\uparrow}\rangle$
($|R^{\downarrow}\rangle$) and $|L^{\uparrow}\rangle$
($|L^{\downarrow}\rangle$) represent the right-circularly polarized
photon and the left-circularly polarized photon propagating along
(against) the normal direction of the cavity Z axis (the
quantization axis), respectively.}\label{fig1}
\end{figure*}

\section{Interaction between a circularly polarized light and a single charged QD in a double-sided microcavity}

A singly charged electron In(Ga)As QD or a GaAs interface QD is
embedded in an optical resonant double-sided microcavity with two
mirrors partially reflective in the top and the bottom, as shown in
Fig.\;\ref{fig1}(a). The optical excitation of a photon and an
excess electron injected into the QD can create an exciton $X^{-}$
consisting of two electrons bound to one heavy hole with negative
charges, as shown in Fig.\;\ref{fig1}(b). If the excess electron in
the QD is in the spin state $|\uparrow\rangle$, only the photon in
the state $|R^{\uparrow}\rangle$ or $|L^{\downarrow}\rangle$ can be
resonantly absorbed to create the exciton state
$|\uparrow\downarrow\Uparrow\rangle$; otherwise, if the excess
electron in the QD is in the spin state $|\downarrow\rangle$, only
the photon in the state $|L^{\uparrow}\rangle$ or
$|R^{\downarrow}\rangle$ can be resonantly absorbed to the exciton
state $|\downarrow\uparrow\Downarrow\rangle$. Here
$|\Uparrow\rangle$ and $|\Downarrow\rangle$ represent the heavy-hole
spin state $|+\frac{3}{2}\rangle$ and $|-\frac{3}{2}\rangle$,
respectively.

The dipole interaction process can be represented by Heisenberg
equations for the cavity field operator $\widehat{a}$ and the
exciton $X^{-}$ dipole operator $\sigma_{-}$ in the interaction
picture \cite{Quant},
\begin{eqnarray}    \label{eq1}
\begin{split}
\frac{d\widehat{a}}{dt}\;=\;&-\left[i(\omega_{c}-\omega)+\kappa+\frac{\kappa_{s}}{2}\right]\widehat{a}
-g\sigma_{-}\\
&-\sqrt{\kappa}\,\widehat{a}_{in'}-\sqrt{\kappa}\,\widehat{a}_{in}+\widehat{H},\\
\frac{d\sigma_{-}}{dt}\;=\;&-\left[i(\omega_{X^{-}}-\omega)+\frac{\gamma}{2}\right]\sigma_{-}
-g\sigma_{z}\,\widehat{a}+\widehat{G},\\
\widehat{a}_{r}\;=\;&\widehat{a}_{in}+\sqrt{\kappa}\,\widehat{a},\;\;\;\;\;\;\;
\widehat{a}_{t}\;=\;\widehat{a}_{in'}+\sqrt{\kappa}\,\widehat{a},
\end{split}
\end{eqnarray}
where $\omega$, $\omega_{c}$, and $\omega_{X^{-}}$ are the
frequencies of the photon, the cavity, and $X^{-}$ transition,
respectively. $g$ is the coupling constant between the $X^{-}$ and
the cavity. $\kappa$ and $\kappa_{s}$ represent the cavity decay
rate and the leaky rate, respectively. $\gamma$ is the exciton
dipole decay rate. $\widehat{H}$ and $\widehat{G}$ are the noise
operators related to reservoirs. $\widehat{a}_{in}$,
$\widehat{a}_{in'}$, $\widehat{a}_{r}$, and $\widehat{a}_{t}$ are
the input and output field operators. In the weak excitation
approximation, the reflection coefficient $r(\omega)$ and the
transmission coefficient $t(\omega)$ of the QD-cavity system can be
described by \cite{QD9,QD10,Inputoutput}
\begin{eqnarray}    \label{eq2}
\begin{split}
r(\omega)\;=\;&1+t(\omega),\\
t(\omega)\;=\;&-\frac{\kappa[i(\omega_{X^{-}}-\omega)+\frac{\gamma}{2}]}{[i(\omega_{X^{-}}-\omega)
+\frac{\gamma}{2}][i(\omega_{c}-\omega)+\kappa+\frac{\kappa_{s}}{2}]+g^{2}}.
\end{split}
\end{eqnarray}
When the QD is uncoupled to the cavity (cold cavity), that is $g=0$,
the reflection $r_{0}(\omega)$ and the transmission $t_{0}(\omega)$
coefficients can be written as
\begin{eqnarray}    \label{eq3}
\begin{split}
r_{0}(\omega)\;=\;&\frac{i(\omega_{c}-\omega)
+\frac{\kappa_{s}}{2}}{i(\omega_{c}-\omega)+\kappa+\frac{\kappa_{s}}{2}},\\
t_{0}(\omega)\;=\;&-\frac{\kappa}{i(\omega_{c}-\omega)+\kappa+\frac{\kappa_{s}}{2}}.
\end{split}
\end{eqnarray}
If $\omega=\omega_{c}=\omega_{X^{-}}$, the reflection and
transmission coefficients of the coupled cavity and the uncoupled
cavity can be simplified as
\begin{eqnarray}    \label{eq4}
\begin{split}
r\;=\;&1+t,\;\;\;\; \;\;\;\;\;\;\;\;
t\;=\;-\frac{\frac{\gamma}{2}\kappa}{\frac{\gamma}{2}(\kappa+\frac{\kappa_{s}}{2})+g^{2}},\\
r_{0}\;=\;&\frac{\frac{\kappa_{s}}{2}}{\kappa+\frac{\kappa_{s}}{2}},\;\;\;\;
\;\;\;\;  t_{0}\;=\;-\frac{\kappa}{\kappa+\frac{\kappa_{s}}{2}}.
\end{split}
\end{eqnarray}
The rules for optical transitions in a realistic QD-cavity system
can be described as \cite{QD10},
\begin{eqnarray}    \label{eq5}
\begin{split}
&|R^{\uparrow}\uparrow\rangle \;\rightarrow
\;r|L^{\downarrow}\uparrow\rangle + t|R^{\uparrow}\uparrow\rangle,
\\
&
|L^{\downarrow}\uparrow\rangle \; \rightarrow \;r|R^{\uparrow}\uparrow\rangle+t|L^{\downarrow}\uparrow\rangle,\\
&|R^{\downarrow}\uparrow\rangle \; \rightarrow \;
t_{0}|R^{\downarrow}\uparrow\rangle
+r_{0}|L^{\uparrow}\uparrow\rangle,\\
&
|L^{\uparrow}\uparrow\rangle \; \rightarrow t_{0}|L^{\uparrow}\uparrow\rangle +r_{0}|R^{\downarrow}\uparrow\rangle,\\
&|R^{\downarrow}\downarrow\rangle \; \rightarrow
r|L^{\uparrow}\downarrow\rangle + t|R^{\downarrow}\downarrow\rangle,
\\
&
|L^{\uparrow}\downarrow\rangle \;\rightarrow \;r|R^{\downarrow}\downarrow\rangle + t |L^{\uparrow}\downarrow\rangle,\\
&|R^{\uparrow}\downarrow\rangle \;\rightarrow\;
t_{0}|R^{\uparrow}\downarrow\rangle+r_{0}|L^{\downarrow}\downarrow\rangle,
\\
& |L^{\downarrow}\downarrow\rangle \;\rightarrow\; t_{0}
|L^{\downarrow}\downarrow\rangle
+r_{0}|R^{\uparrow}\downarrow\rangle.
\end{split}
\end{eqnarray}

\section{An error-detected block for the interaction between the circularly-polarized photon and the QD-cavity system}

\begin{figure}[th]
\centering
\includegraphics[width=6cm,angle=0]{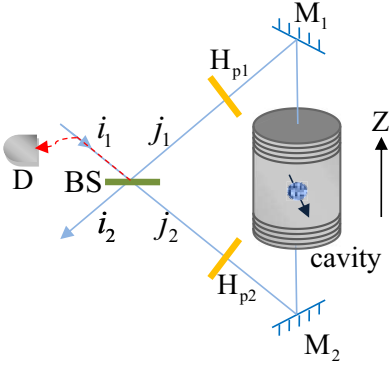}
\caption{ Schematic diagram of the error-detected  block. BS is a
$50:50$ beam splitter which performs the  spatial-mode Hadamard
operation
[$|i_{1}\rangle\rightarrow\frac{1}{\sqrt{2}}(|j_{1}\rangle+|j_{2}\rangle)$,
$|i_{2}\rangle\rightarrow\frac{1}{\sqrt{2}}(|j_{1}\rangle-|j_{2}\rangle)$]
on  the photon. H$_{pi}$ $(i=1,2)$ is a half-wave plate which
performs the polarization Hadamard operation
[$|R\rangle\rightarrow\frac{1}{\sqrt{2}}(|R\rangle+|L\rangle)$,
$|L\rangle\rightarrow\frac{1}{\sqrt{2}}(|R\rangle-|L\rangle)$] on
the   photon. M$_{i}$ $(i=1,2)$ is a mirror. D is a single-photon
detector.}  \label{fig2}
\end{figure}

The schematic diagram of our error-detected block for the
interaction between the circularly-polarized photon and the
QD-cavity system is shown in Fig.\;\ref{fig2}, which is constructed
with a $50:50$ beam splitter (BS), two half-wave plates (H$_{pi}$),
two mirrors (M$_{i}$), a single-photon detector (D), and a QD-cavity
system. The QD in the cavity is prepared in the state
$|\varphi_{+}\rangle=\frac{1}{\sqrt{2}}(|\uparrow\rangle+|\downarrow\rangle)$.
If the photon is in the right-circularly polarized state
$|R\rangle$, one injects it into our error-detected block from the
path $i_{1}$ and lets it pass through BS and H$_{pi}$ $(i=1,2)$.
After being reflected by the mirror M$_{i}$ $(i=1,2)$, the state of
the whole system composed of the photon and the QD in the cavity is
changed from the state $|\Phi\rangle_{0}=|R\rangle|i_{1}\rangle
\otimes \frac{1}{\sqrt{2}}(|\uparrow\rangle+|\downarrow\rangle)$
into the state $|\Phi\rangle_{1}$. Here  $|\Phi\rangle_{1}$ is
\begin{eqnarray}    \label{eq6}
\begin{split}
|\Phi\rangle_{1}\;=\;&\frac{1}{2}[(|R^{\downarrow}\rangle
+|L^{\downarrow}\rangle)|j_{1}\rangle
+(|R^{\uparrow}\rangle+|L^{\uparrow}\rangle)|j_{2}\rangle]\\&
\otimes\frac{1}{\sqrt{2}}(|\uparrow\rangle+|\downarrow\rangle).
\end{split}
\end{eqnarray}
After the interaction between the photon and the QD-cavity system,
the state becomes
\begin{eqnarray}    \label{eq7}
\begin{split}
|\Phi\rangle_{2}\;=\;&\frac{1}{2\sqrt{2}}[(t_{0}+r_{0})|R^{\downarrow}\rangle|j_{2}\rangle
+(t+r)|L^{\downarrow}\rangle|j_{2}\rangle\\
&+(t+r)|R^{\uparrow}\rangle|j_{1}\rangle+(r_{0}+t_{0})|L^{\uparrow}\rangle|j_{1}\rangle]|\uparrow\rangle
\\
&+\frac{1}{2\sqrt{2}}[(t_{0}+r_{0})|R^{\uparrow}\rangle|j_{1}\rangle
+(t+r)|L^{\uparrow}\rangle|j_{1}\rangle
\\
&+(t+r)|R^{\downarrow}\rangle|j_{2}\rangle
+(t_{0}+r_{0})|L^{\downarrow}\rangle|j_{2}\rangle]|\downarrow\rangle.
\end{split}
\end{eqnarray}
Subsequently, the photon is reflected by M$_{i}$ $(i=1,2)$ and
passes through H$_{pi}$ $(i=1,2)$ again, and the state evolves to
\begin{eqnarray}    \label{eq8}
\begin{split}
|\Phi\rangle_{3}\;=\;&\frac{1}{4}[(t+r+t_{0}+r_{0})|R^{\downarrow}\rangle|j_{2}\rangle
\\
&+(t+r+t_{0}+r_{0})|R^{\uparrow}\rangle|j_{1}\rangle \\
&-(t+r-t_{0}-r_{0})|L^{\downarrow}\rangle|j_{2}\rangle
\\
&+(t+r-t_{0}-r_{0})|L^{\uparrow}\rangle|j_{1}\rangle]|\uparrow\rangle \\
&+\frac{1}{4}[(t+r+t_{0}+r_{0})|R^{\downarrow}\rangle|j_{2}\rangle
\\
&+(t+r+t_{0}+r_{0})|R^{\uparrow}\rangle|j_{1}\rangle \\
&+(t+r-t_{0}-r_{0})|L^{\downarrow}\rangle|j_{2}\rangle
\\
&-(t+r-t_{0}-r_{0})|L^{\uparrow}\rangle|j_{1}\rangle]|\downarrow\rangle.
\end{split}
\end{eqnarray}
At last, the photon passes through BS again and the final state can
be described as
\begin{eqnarray}    \label{eq9}
|\Phi\rangle_{4}=D|R\rangle|i_{1}\rangle|\varphi_{+}\rangle+T|L\rangle|i_{2}\rangle|\varphi_{-}\rangle.
\end{eqnarray}
Here $D=\frac{1}{2}(t+r+t_{0}+r_{0})$ and
$T=\frac{1}{2}(t+r-t_{0}-r_{0})$ are the reflection coefficient and
the transmission coefficient of the error-detected block,
respectively. Similarly, if the QD in the cavity is prepared in the
state
$|\varphi_{-}\rangle=\frac{1}{\sqrt{2}}(|\uparrow\rangle-|\downarrow\rangle)$,
the outcome of the process can be described as
\begin{eqnarray}    \label{eq10}
|\Phi\rangle_{5}=D|R\rangle|i_{1}\rangle|\varphi_{-}\rangle+T|L\rangle|i_{2}\rangle|\varphi_{+}\rangle.
\end{eqnarray}

\begin{figure}[th]
\centering
\includegraphics[width=6.5cm,angle=0]{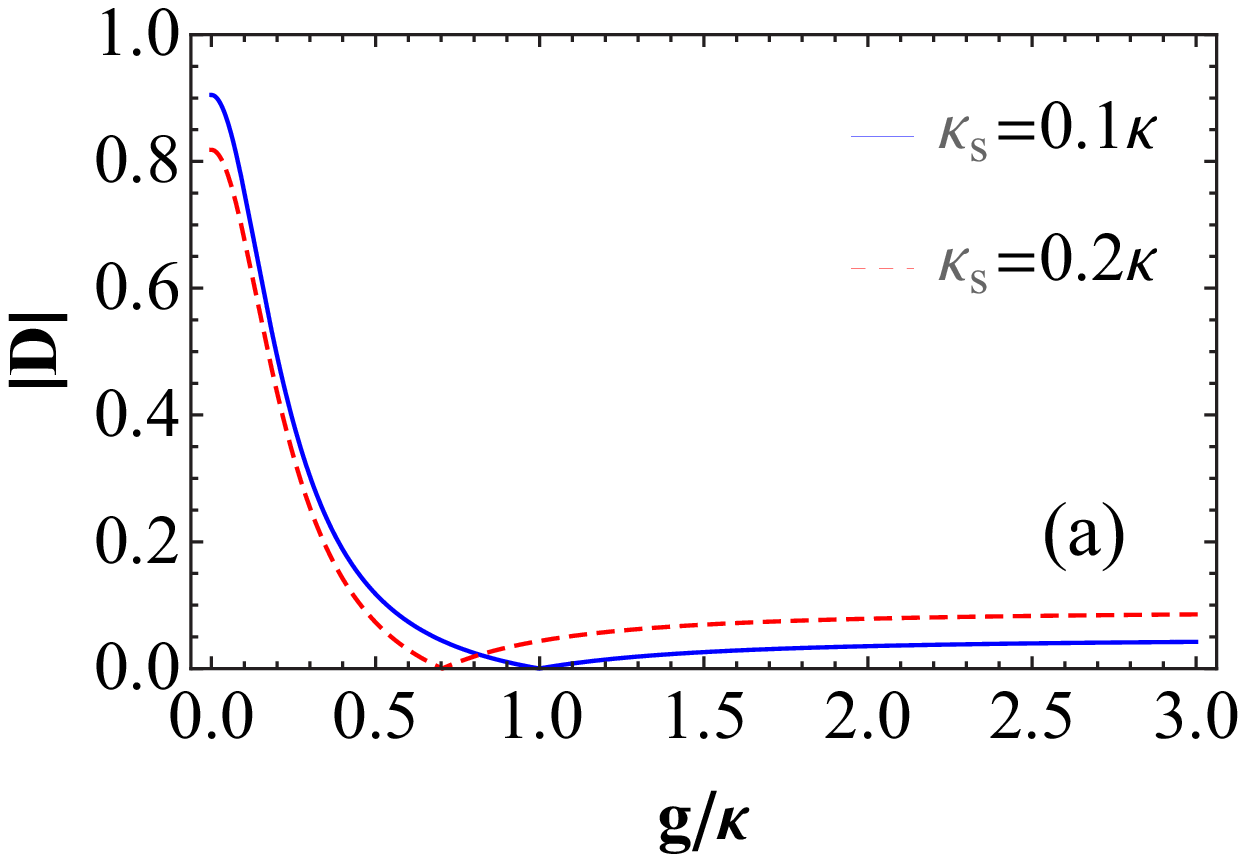} \hspace{20pt}
\includegraphics[width=6.5cm,angle=0]{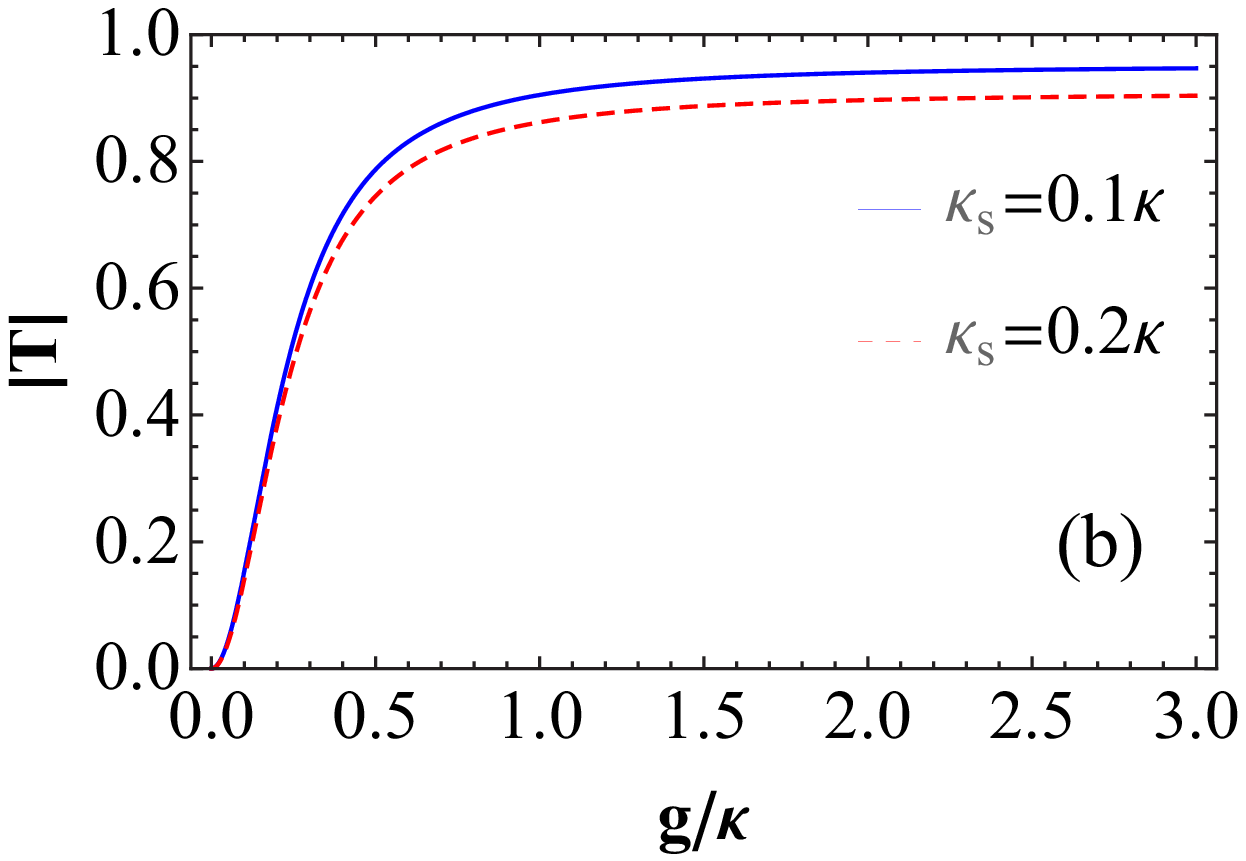}
\caption{(a) The blue solid line and the red dashed line are the
reflection coefficient $|D|$ of the error-detected block vs the
normalized coupling strength $g/\kappa$ for the leakage rates
$\kappa_{s}=0.1\kappa$ and $\kappa_{s}=0.2\kappa$, respectively. (b)
The blue solid line and the red dashed line are the transmission
coefficient $|T|$ of the error-heralded block vs $g/\kappa$ for
$\kappa_{s}=0.1\kappa$ and $\kappa_{s}=0.2\kappa$, respectively.
$\gamma=0.1\kappa$, which is experimentally achievable, and
$\omega=\omega_{c}=\omega_{X^{-}}$ are taken here.} \label{fig3}
\end{figure}

One can see that there are two parts of the outcome. If the photon
is reflected from the error-detected block with probability of
$|D|^{2}$, the polarization of the photon and the state of the QD
would not change. The reflected photon would be detected by the
detector and the click of the detector represents the case in which
the corresponding task of the error-detected block runs fail. If the
photon is transmitted from the error-detected block with probability
of $|T|^{2}$, the polarization of the photon is flipped and the
superposition state of the QD is changed from $|\varphi_{+}\rangle$
to $|\varphi_{-}\rangle$ or from $|\varphi_{-}\rangle$ to
$|\varphi_{+}\rangle$. We can utilize the transmitted photon and the
QD-cavity to accomplish the  tasks of deterministic HBSG, complete
HBSA, and complete nondestructive HBSA with a high fidelity. $|D|$
and $|T|$ are affected by the $g/\kappa$ and $\kappa_{s}/\kappa$ as
shown in Fig.\;\ref{fig3}. From Fig.\;\ref{fig3}, one can see that
there exists a zero value of reflection coefficient, which results
from the destructive interference.  For the condition
$g^{2}/\kappa\gamma\gg1$ and $\kappa_{s}/2\kappa\gg1$, the final
state $|\Phi\rangle_{f1}=|L\rangle|i_{2}\rangle|\varphi_{-}\rangle$
and $|\Phi\rangle_{f2}=|L\rangle|i_{2}\rangle|\varphi_{+}\rangle$ of
the error-detected block is obtained when the QD is prepared in the
state $|\varphi_{+}\rangle$ and $|\varphi_{-}\rangle$, respectively.

\section{Deterministic photonic hyperentanglement generation}

A two-photon four-qubit hyperentangled Bell state can be described
as
\begin{eqnarray}    \label{eq11}
|\Psi\rangle_{PS}=\frac{1}{2}(|RR\rangle+|LL\rangle)_{AB}(|a_{1}b_{1}\rangle
+|a_{2}b_{2}\rangle)_{AB}.
\end{eqnarray}
Here the subscripts $P$ and $S$ denote the polarization and the
spatial-mode DOFs, respectively. The subscripts $A$ and $B$
represent the two photons in the hyperentangled state, respectively.
$|R\rangle$ and $|L\rangle$ represent the right-circular and the
left-circular polarizations of photons, respectively.
$|a_{1}\rangle$ ($|b_{1}\rangle$) and $|a_{2}\rangle$
($|b_{2}\rangle$) are the different spatial modes for photon $A$
($B$). The four Bell states in the polarization DOF can be expressed
as
\begin{eqnarray}    \label{eq12}
\begin{split}
|\phi^{\pm}\rangle_{P}\;=\;&\frac{1}{\sqrt{2}}(|RR\rangle \pm
|LL\rangle),\\
|\psi^{\pm}\rangle_{P}\;=\;& \frac{1}{\sqrt{2}}(|RL\rangle \pm
|LR\rangle).
\end{split}
\end{eqnarray}
The four Bell states in the spatial-mode DOF can be written as
\begin{eqnarray}    \label{eq13}
\begin{split}
|\phi^{\pm}\rangle_{S}\;=\;&\frac{1}{\sqrt{2}}(|a_{1}b_{1}\rangle
\pm
|a_{2}b_{2}\rangle),\\
|\psi^{\pm}\rangle_{S}\;=\;& \frac{1}{\sqrt{2}}(|a_{1}b_{2}\rangle
\pm |a_{2}b_{1}\rangle).
\end{split}
\end{eqnarray}

\begin{figure*}[th]
\centering
\includegraphics[width=14cm,angle=0]{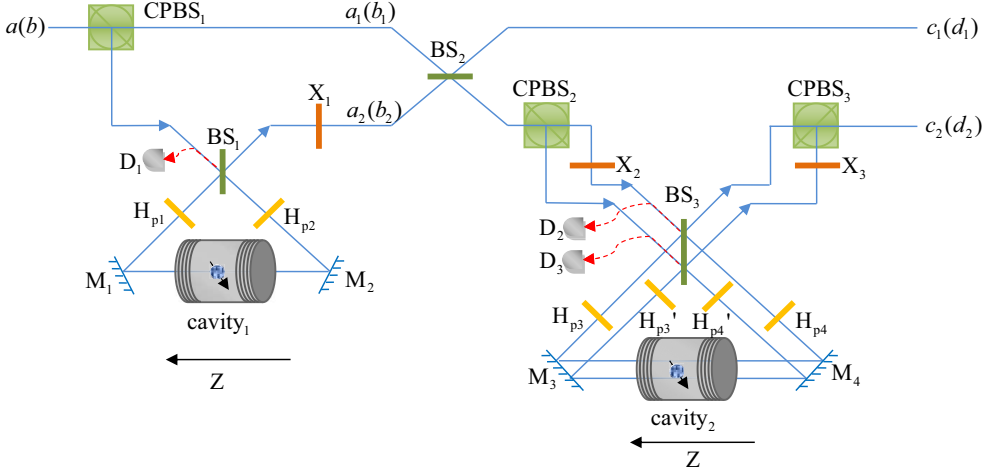}
\caption{ Schematic diagram for two-photon polarization-spatial
HBSG. CPBS$_{i}$ $(i=1,2,3)$ is a circularly polarized beam splitter
which transmits the photon in the left-circular polarization
$|L\rangle$ and reflects the photon in the right-circular
polarization $|R\rangle$, respectively. X$_{i}$ $(i=1,2,3)$ is a
half-wave plate which performs a polarization bit-flip operation
$\sigma_{x}^{p}=|R\rangle \langle L|+|L\rangle \langle R|$ on the
photon.}  \label{fig4}
\end{figure*}

The principle of our scheme for the two-photon polarization-spatial
hyperentangled-Bell-state generation (HBSG) constructed with two
error-detected blocks and some linear optical elements is shown in
Fig.\;\ref{fig4}. The QDs are prepared in the initial states
$|\varphi_{+}\rangle_{1}=|\varphi_{+}\rangle_{2}$ and the two
photons $A$ and $B$ with the same frequency are prepared in the same
initial state
$|\varphi\rangle_{A}=|\varphi\rangle_{B}=\frac{1}{\sqrt{2}}(|R\rangle+|L\rangle)$.
The process for HBSG can be described in detail as follows.

First, one injects  photon $A$ into the quantum circuit from the
left input, followed by photon $B$. The time interval $\Delta t$
exists between  photons $A$ and $B$, and $\Delta t$ should be less
than the spin coherence time $\Gamma$. CPBS$_{1}$ will transmit the
photon to path $a_{1}$ $(b_{1})$ and reflects the photon to path
$a_{2}$ $(b_{2})$ when  photon $A$ $(B)$ is in the state $|L\rangle$
and $|R\rangle$, respectively. When  photon $A$ $(B)$ emerges in
path $a_{2}$ $(b_{2})$, it passes through the error-detected block
consisting of the QD-cavity$_{1}$ system and X$_{1}$. Before the two
wave packets split by CPBS$_{1}$  interfere  at BS$_{2}$, the state
of the whole system is changed from $\Omega_{0}=|\varphi\rangle_{A}
|\varphi\rangle_{B} |\varphi_{+}\rangle_{1} |\varphi_{+}\rangle_{2}$
to $\Omega_{1}$. Here  $\Omega_{1}$ is written as
\begin{eqnarray}    \label{eq14}
\begin{split}
\Omega_{1}=&\frac{1}{2}(|LL\rangle |a_{1}b_{1}\rangle |\varphi_{+}\rangle_{1}+ |LR\rangle |a_{1}b_{2}\rangle |\varphi_{-}\rangle_{1} \\
&+ |RL\rangle |a_{2}b_{1}\rangle |\varphi_{-}\rangle_{1} +|RR\rangle
|a_{2}b_{2}\rangle |\varphi_{+}\rangle_{1}) |\varphi_{+}\rangle_{2}.
\end{split}
\end{eqnarray}
Then BS$_{2}$, which performs the spatial-mode Hadamard operation
\big[$|a_{1}(b_{1})\rangle\rightarrow
\frac{1}{\sqrt{2}}(|c_{1}(|d_{1})\rangle+|c_{2}(d_{2})\rangle$,
$|a_{2}(b_{2})\rangle\rightarrow
\frac{1}{\sqrt{2}}(|c_{1}(|d_{1})\rangle-|c_{2}(d_{2})\rangle$\big]
on photons transmitted, will lead photon $A$ $(B)$ to  paths $c_{1}$
$(d_{1})$ and $c_{2}$ $(d_{2})$. In path $c_{2}$ $(d_{2})$,  photon
$A$ $(B)$ passes through CPBS$_{2}$ which transmits  the photon $A$
$(B)$ in $|L\rangle$ and reflects the photon in $|R\rangle$,
respectively. Then  photon $A$ $(B)$ in $|L\rangle$ takes a
$\sigma_{x}^{p}$ operation by X$_{2}$ and passes through the
error-detected block consisting of QD-cavity$_{2}$ system,
sequently. Photon $A$ $(B)$ in $|R\rangle$ passes through the same
error-detected block and takes a $\sigma_{x}^{p}$ operation by
X$_{3}$, sequentially. At last, the two wave packets union at
CPBS$_{3}$ in path $a_{2}$ $(b_{2})$. The state of the whole system
becomes
\begin{eqnarray}    \label{eq15}
\begin{split}
\Omega_{2}\;=\;&\frac{1}{2}(|\phi^{+}\rangle_{P}|\phi^{+}\rangle_{S}|\varphi_{+}\rangle_{1}|\varphi_{+}\rangle_{2}\\
&
-|\phi^{-}\rangle_{P}|\psi^{+}\rangle_{S}|\varphi_{+}\rangle_{1}|\varphi_{-}\rangle_{2}\\
&+|\psi^{+}\rangle_{P}|\phi^{-}\rangle_{S}|\varphi_{-}\rangle_{1}|\varphi_{+}\rangle_{2}\\
&
+|\psi^{-}\rangle_{P}|\psi^{-}\rangle_{S}|\varphi_{-}\rangle_{1}|\varphi_{-}\rangle_{2}).
\end{split}
\end{eqnarray}
From Eq.\;(\ref{eq15}), one can see the relationship between the
measurement outcomes of the two QD-cavity systems and the
polarization-spatial hyperentangled Bell states of the two-photon
system. If QD$_{1}$ and QD$_{2}$ are in the states
$|\varphi_{+}\rangle_{1}$ and  $|\varphi_{+}\rangle_{2}$,
respectively, the two-photon system is in the hyperentangled Bell
state $|\phi^{+}\rangle_{P}|\phi^{+}\rangle_{S}$. If QD$_{1}$ and
QD$_{2}$ are in the states $|\varphi_{+}\rangle_{1}$ and
$|\varphi_{-}\rangle_{2}$, respectively, the two-photon system is in
the hyperentangled Bell state
$|\phi^{-}\rangle_{P}|\psi^{+}\rangle_{S}$. When QD$_{1}$ and
QD$_{2}$ are in the states $|\varphi_{-}\rangle_{1}$ and
$|\varphi_{+}\rangle_{2}$, respectively, the two-photon system is in
the hyperentangled Bell state
$|\psi^{+}\rangle_{P}|\phi^{-}\rangle_{S}$. When QD$_{1}$ and
QD$_{2}$ are in the states $|\varphi_{-}\rangle_{1}$ and
$|\varphi_{-}\rangle_{2}$, respectively, the two-photon system is in
the hyperentangled Bell state
$|\psi^{-}\rangle_{P}|\psi^{-}\rangle_{S}$. Therefore, one can
generate deterministically a polarization-spatial hyperentangled
Bell state of the two-photon system by measuring the states of the
two QDs. Other polarization-spatial hyperentangled Bell states can
be generated with some appropriate single-qubit operations.

\begin{figure*}[th]
\centering
\includegraphics[width=15cm,angle=0]{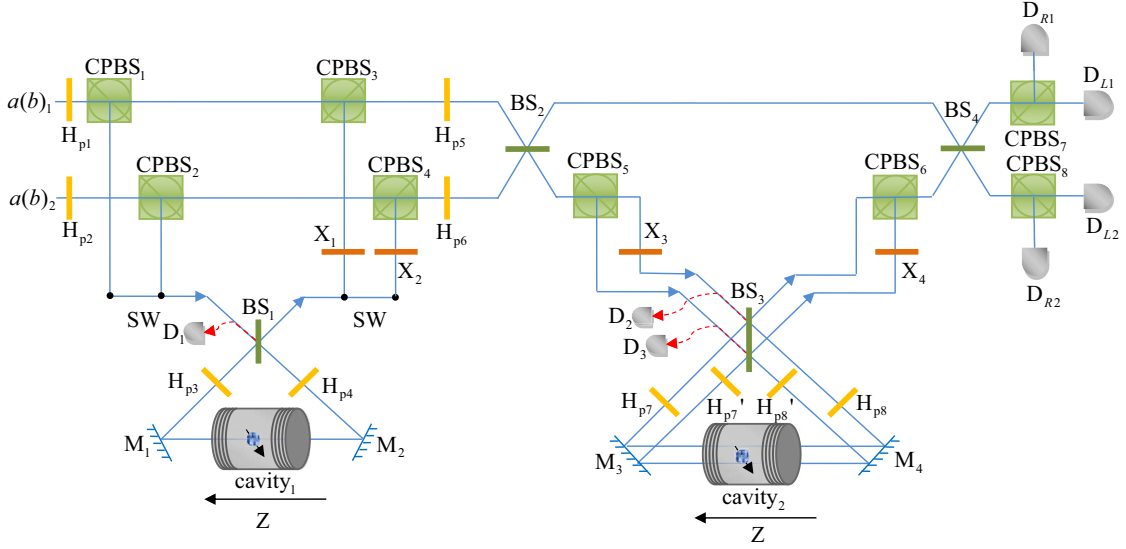}
\caption{ Schematic diagram of the complete polarization-spatial
HBSA. D$_{i}$ $(i=1,2,3)$, D$_{Li}$ $(i=1,2)$ and D$_{Ri}$ $(i=1,2)$
are single-photon detectors. SW is an optical switch which lets the
wave-packets of a photon in different spatial-mode pass into and out
of the error-detected block sequentially.}  \label{fig5}
\end{figure*}

\section{Complete polarization-spatial hyperentangled Bell states analysis}

The schematic diagram of the complete polarization-spatial HBSA for
hyperentangled two-photon systems is shown in Fig.\;\ref{fig5}. Here
we use two error-detected blocks consisting of two QD-cavity systems
and some linear optical elements to achieve the complete
polarization-spatial HBSA. The two QDs are prepared in the initial
states $|\varphi_{+}\rangle_{1}=|\varphi_{+}\rangle_{2}$ and the
hyperentangled two-photon system is in one of the $16$
hyperentangled Bell states. The process for our HBSA can be
described as follows.

One lets photon $A$ pass through the quantum circuit from the left
input port, followed by photon $B$. And the two photons will be
detected sequentially at the output port. The interval time $\Delta
t$ existing between the two photons is less than the spin coherence
time $\Gamma$. According to the results of the interaction between
the circularly polarized photon and the error-detected block, after
the two photons $A$ and $B$ pass through quantum circuit and before
they are detected by the detectors (D$_{Ri}$ and D$_{Li}$), the
whole system composed of the two photons and the two QDs evolves as
\begin{widetext}
\begin{eqnarray}    \label{eq16}
\begin{split}
&|\phi^{+}\rangle_{P}(|\psi^{+}\rangle_{P})|\phi^{+}\rangle_{S}
|\varphi_{+}\rangle_{1}|\varphi_{+}\rangle_{2}
\;\;\,\rightarrow \;\; |\phi^{+}\rangle_{P}(|\psi^{+}\rangle_{P})|\phi^{+}\rangle_{S} |\varphi_{+}\rangle_{1}|\varphi_{+}\rangle_{2},\\
&|\phi^{+}\rangle_{P}(|\psi^{+}\rangle_{P})|\psi^{+}\rangle_{S}
|\varphi_{+}\rangle_{1}|\varphi_{+}\rangle_{2}
\;\;\rightarrow \;\; |\phi^{+}\rangle_{P}(|\psi^{+}\rangle_{P})|\psi^{+}\rangle_{S} |\varphi_{+}\rangle_{1}|\varphi_{+}\rangle_{2},\\
&|\phi^{+}\rangle_{P}(|\psi^{+}\rangle_{P})|\phi^{-}\rangle_{S}
|\varphi_{+}\rangle_{1}|\varphi_{+}\rangle_{2}
\;\;\,\rightarrow \;\; |\phi^{+}\rangle_{P}(|\psi^{+}\rangle_{P})|\phi^{-}\rangle_{S} |\varphi_{+}\rangle_{1}|\varphi_{-}\rangle_{2},\\
&|\phi^{+}\rangle_{P}(|\psi^{+}\rangle_{P})|\psi^{-}\rangle_{S}
|\varphi_{+}\rangle_{1}|\varphi_{+}\rangle_{2}
\;\;\rightarrow \;\; |\phi^{+}\rangle_{P}(|\psi^{+}\rangle_{P})|\psi^{-}\rangle_{S} |\varphi_{+}\rangle_{1}|\varphi_{-}\rangle_{2},\\
&|\phi^{-}\rangle_{P}(|\psi^{-}\rangle_{P})|\phi^{+}\rangle_{S}
|\varphi_{+}\rangle_{1}|\varphi_{+}\rangle_{2}
\;\;\,\rightarrow \;\; |\phi^{-}\rangle_{P}(|\psi^{-}\rangle_{P})|\phi^{+}\rangle_{S} |\varphi_{-}\rangle_{1}|\varphi_{+}\rangle_{2},\\
&|\phi^{-}\rangle_{P}(|\psi^{-}\rangle_{P})|\psi^{+}\rangle_{S}
|\varphi_{+}\rangle_{1}|\varphi_{+}\rangle_{2}
\;\;\rightarrow \;\; |\phi^{-}\rangle_{P}(|\psi^{-}\rangle_{P})|\psi^{+}\rangle_{S} |\varphi_{-}\rangle_{1}|\varphi_{+}\rangle_{2},\\
&|\phi^{-}\rangle_{P}(|\psi^{-}\rangle_{P})|\phi^{-}\rangle_{S}
|\varphi_{+}\rangle_{1}|\varphi_{+}\rangle_{2}
\;\;\,\rightarrow \;\; |\phi^{-}\rangle_{P}(|\psi^{-}\rangle_{P})|\phi^{-}\rangle_{S} |\varphi_{-}\rangle_{1}|\varphi_{-}\rangle_{2},\\
&|\phi^{-}\rangle_{P}(|\psi^{-}\rangle_{P})|\psi^{-}\rangle_{S}
|\varphi_{+}\rangle_{1}|\varphi_{+}\rangle_{2} \;\; \rightarrow \;\;
|\phi^{-}\rangle_{P}(|\psi^{-}\rangle_{P})|\psi^{-}\rangle_{S}
|\varphi_{-}\rangle_{1}|\varphi_{-}\rangle_{2}.
\end{split}
\end{eqnarray}
\end{widetext}

At last, the photons $A$ and $B$ are independently measured in both
the polarization and the spatial-mode DOFs with single-photon
detectors, and the two QDs are measured in the basis
$\{|\varphi_{+}\rangle, |\varphi_{-}\rangle\}$. The relationship
between the measurement outcomes and the initial hyperentangled
states of the two-photon system $AB$ is shown in
Table\;\ref{table1}.


\begin{table}[htb]
\centering \caption{ The relationship between the measurement
outcomes of the states of the two-photon systems in the polarization
and spatial-mode DOFs and the two QDs and the initial hyperentangled
states of the two-photon system $|\Psi\rangle_{PS}$.}
\begin{tabular}{cccc}
\hline\hline
QD$_{1}$ and QD$_{2}$                           &     Polarization    &   Spatial-mode              & $|\Psi\rangle_{PS}$      \\
   \hline
$|\varphi_{+}\rangle_{1}|\varphi_{+}\rangle_{2}$   &    $RR$, $LL$ &
$a_{1}b_{1}$, $a_{2}b_{2}$  &
$|\phi^{+}\rangle_{P}|\phi^{+}\rangle_{S}$\\ 
$|\varphi_{+}\rangle_{1}|\varphi_{+}\rangle_{2}$   &    $RR$, $LL$ &
$a_{1}b_{2}$, $a_{2}b_{1}$  &
$|\phi^{+}\rangle_{P}|\psi^{+}\rangle_{S}$\\ 
$|\varphi_{+}\rangle_{1}|\varphi_{+}\rangle_{2}$   &    $RL$, $LR$ &
$a_{1}b_{1}$, $a_{2}b_{2}$  &
$|\psi^{+}\rangle_{P}|\phi^{+}\rangle_{S}$\\ 
$|\varphi_{+}\rangle_{1}|\varphi_{+}\rangle_{2}$   &    $RL$, $LR$ &
$a_{1}b_{2}$, $a_{2}b_{1}$  &
$|\psi^{+}\rangle_{P}|\psi^{+}\rangle_{S}$\\ 
$|\varphi_{+}\rangle_{1}|\varphi_{-}\rangle_{2}$   &    $RR$, $LL$ &
$a_{1}b_{1}$, $a_{2}b_{2}$  &
$|\phi^{+}\rangle_{P}|\phi^{-}\rangle_{S}$\\ 
$|\varphi_{+}\rangle_{1}|\varphi_{-}\rangle_{2}$   &    $RR$, $LL$ &
$a_{1}b_{2}$, $a_{2}b_{1}$  &
$|\phi^{+}\rangle_{P}|\psi^{-}\rangle_{S}$\\ 
$|\varphi_{+}\rangle_{1}|\varphi_{-}\rangle_{2}$   &    $RL$, $LR$ &
$a_{1}b_{1}$, $a_{2}b_{2}$  &
$|\psi^{+}\rangle_{P}|\phi^{-}\rangle_{S}$\\ 
$|\varphi_{+}\rangle_{1}|\varphi_{-}\rangle_{2}$   &    $RL$, $LR$ &
$a_{1}b_{2}$, $a_{2}b_{1}$  &
$|\psi^{+}\rangle_{P}|\psi^{-}\rangle_{S}$\\ 
$|\varphi_{-}\rangle_{1}|\varphi_{+}\rangle_{2}$   &    $RR$, $LL$ &
$a_{1}b_{1}$, $a_{2}b_{2}$  &
$|\phi^{-}\rangle_{P}|\phi^{+}\rangle_{S}$\\ 
$|\varphi_{-}\rangle_{1}|\varphi_{+}\rangle_{2}$   &    $RR$, $LL$ &
$a_{1}b_{2}$, $a_{2}b_{1}$  &
$|\phi^{-}\rangle_{P}|\psi^{+}\rangle_{S}$\\ 
$|\varphi_{-}\rangle_{1}|\varphi_{+}\rangle_{2}$   &    $RL$, $LR$ &
$a_{1}b_{1}$, $a_{2}b_{2}$  &
$|\psi^{-}\rangle_{P}|\phi^{+}\rangle_{S}$\\ 
$|\varphi_{-}\rangle_{1}|\varphi_{+}\rangle_{2}$   &    $RL$, $LR$ &
$a_{1}b_{2}$, $a_{2}b_{1}$  &
$|\psi^{-}\rangle_{P}|\psi^{+}\rangle_{S}$\\ 
$|\varphi_{-}\rangle_{1}|\varphi_{-}\rangle_{2}$   &    $RR$, $LL$ &
$a_{1}b_{1}$, $a_{2}b_{2}$  &
$|\phi^{-}\rangle_{P}|\phi^{-}\rangle_{S}$\\ 
$|\varphi_{-}\rangle_{1}|\varphi_{-}\rangle_{2}$   &    $RR$, $LL$ &
$a_{1}b_{2}$, $a_{2}b_{1}$  &
$|\phi^{-}\rangle_{P}|\psi^{-}\rangle_{S}$\\ 
$|\varphi_{-}\rangle_{1}|\varphi_{-}\rangle_{2}$   &    $RL$, $LR$ &
$a_{1}b_{1}$, $a_{2}b_{2}$  &
$|\psi^{-}\rangle_{P}|\phi^{-}\rangle_{S}$\\ 
$|\varphi_{-}\rangle_{1}|\varphi_{-}\rangle_{2}$   &    $RL$, $LR$       & $a_{1}b_{2}$, $a_{2}b_{1}$  & $|\psi^{-}\rangle_{P}|\psi^{-}\rangle_{S}$\\
\hline\hline
\end{tabular}\label{table1}
\end{table}

From Table\;\ref{table1}, one can obtain the complete and
deterministic analysis on hyperentangled Bell states of a two-photon
system $AB$. The measurement outcomes of QD$_{1}$ and QD$_{2}$
reveal the phase information in the polarization  and the
spatial-mode DOFs, respectively. In detail, when QD$_{1}$ (QD$_{2})$
is in the spin state $|\varphi_{+}\rangle_{1}$
$(|\varphi_{+}\rangle_{2})$, the two-photon system is in the states
$|\phi^{+}\rangle_{P(S)}$ or $|\psi^{+}\rangle_{P(S)}$ in the
polarization (spatial-mode) DOF. Otherwise, when QD$_{1}$ (QD$_{2})$
is in the spin state $|\varphi_{-}\rangle_{1}$
$(|\varphi_{-}\rangle_{2})$, the two-photon system is in the states
$|\phi^{-}\rangle_{P(S)}$ or $|\psi^{-}\rangle_{P(S)}$ in the
polarization (spatial-mode) DOF. The measurement outcomes of the
states of the two-photon system in the polarization DOF reveal the
parity information in the polarization DOF. In detail, if the
measurement outcome of the state is $RR$ or $LL$, the two-photon
system is in the state $|\phi^{\pm}\rangle_{P}$ in the polarization
DOF; otherwise, it is in the state $|\psi^{\pm}\rangle_{P}$. The
measurement outcomes of the states of the two-photon system in the
spatial-mode DOF reveal the parity information in the spatial-mode
DOF. In detail, if the measurement outcome of the state is
$a_{1}b_{1}$ or $a_{2}b_{2}$, the two-photon system is in the state
$|\phi^{\pm}\rangle_{S}$; otherwise, it is in the state
$|\psi^{\pm}\rangle_{S}$. That is, the schematic diagram shown in
Fig.\;\ref{fig5} can be used for the complete and deterministic
HBSA.

\begin{table}[htb]
\centering \caption{ The relationship between the measurement
outcomes of the states of the four QDs and the initial
hyperentangled states of the two-photon system $|\Psi\rangle_{PS}$.}
\begin{tabular}{ccccc}
\hline\hline
QD$_{1}$                   &QD$_{2}$                    &QD$_{3}$                &QD$_{4}$      & $|\Psi\rangle_{PS}$      \\
\hline $|\varphi_{+}\rangle_{1}$     & $|\varphi_{+}\rangle_{2}$   &
$|\varphi_{+}\rangle_{3}$  & $|\varphi_{+}\rangle_{4}$
&$|\phi^{+}\rangle_{P}|\phi^{+}\rangle_{S}$\\ 
$|\varphi_{-}\rangle_{1}$     & $|\varphi_{+}\rangle_{2}$   &
$|\varphi_{+}\rangle_{3}$  & $|\varphi_{+}\rangle_{4}$
&$|\psi^{+}\rangle_{P}|\phi^{+}\rangle_{S}$\\ 
$|\varphi_{+}\rangle_{1}$     & $|\varphi_{-}\rangle_{2}$   &
$|\varphi_{+}\rangle_{3}$  & $|\varphi_{+}\rangle_{4}$
&$|\phi^{-}\rangle_{P}|\phi^{+}\rangle_{S}$\\ 
$|\varphi_{-}\rangle_{1}$     & $|\varphi_{-}\rangle_{2}$   &
$|\varphi_{+}\rangle_{3}$  & $|\varphi_{+}\rangle_{4}$
&$|\psi^{-}\rangle_{P}|\phi^{+}\rangle_{S}$\\ 
$|\varphi_{+}\rangle_{1}$     & $|\varphi_{+}\rangle_{2}$   &
$|\varphi_{-}\rangle_{3}$  & $|\varphi_{+}\rangle_{4}$
&$|\phi^{+}\rangle_{P}|\psi^{+}\rangle_{S}$\\ 
$|\varphi_{-}\rangle_{1}$     & $|\varphi_{+}\rangle_{2}$   &
$|\varphi_{-}\rangle_{3}$  & $|\varphi_{+}\rangle_{4}$
&$|\psi^{+}\rangle_{P}|\psi^{+}\rangle_{S}$\\ 
$|\varphi_{+}\rangle_{1}$     & $|\varphi_{-}\rangle_{2}$   &
$|\varphi_{-}\rangle_{3}$  & $|\varphi_{+}\rangle_{4}$
&$|\phi^{-}\rangle_{P}|\psi^{+}\rangle_{S}$\\ 
$|\varphi_{-}\rangle_{1}$     & $|\varphi_{-}\rangle_{2}$   &
$|\varphi_{-}\rangle_{3}$  & $|\varphi_{+}\rangle_{4}$
&$|\psi^{-}\rangle_{P}|\psi^{+}\rangle_{S}$\\ 
$|\varphi_{+}\rangle_{1}$     & $|\varphi_{+}\rangle_{2}$   &
$|\varphi_{+}\rangle_{3}$  & $|\varphi_{-}\rangle_{4}$
&$|\phi^{+}\rangle_{P}|\phi^{-}\rangle_{S}$\\ 
$|\varphi_{-}\rangle_{1}$     & $|\varphi_{+}\rangle_{2}$   &
$|\varphi_{+}\rangle_{3}$  & $|\varphi_{-}\rangle_{4}$
&$|\psi^{+}\rangle_{P}|\phi^{-}\rangle_{S}$\\ 
$|\varphi_{+}\rangle_{1}$     & $|\varphi_{-}\rangle_{2}$   &
$|\varphi_{+}\rangle_{3}$  & $|\varphi_{-}\rangle_{4}$
&$|\phi^{-}\rangle_{P}|\phi^{-}\rangle_{S}$\\ 
$|\varphi_{-}\rangle_{1}$     & $|\varphi_{-}\rangle_{2}$   &
$|\varphi_{+}\rangle_{3}$  & $|\varphi_{-}\rangle_{4}$
&$|\psi^{-}\rangle_{P}|\phi^{-}\rangle_{S}$\\ 
$|\varphi_{+}\rangle_{1}$     & $|\varphi_{+}\rangle_{2}$   &
$|\varphi_{-}\rangle_{3}$  & $|\varphi_{-}\rangle_{4}$
&$|\phi^{+}\rangle_{P}|\psi^{-}\rangle_{S}$\\ 
$|\varphi_{-}\rangle_{1}$     & $|\varphi_{+}\rangle_{2}$   &
$|\varphi_{-}\rangle_{3}$  & $|\varphi_{-}\rangle_{4}$
&$|\psi^{+}\rangle_{P}|\psi^{-}\rangle_{S}$\\ 
$|\varphi_{+}\rangle_{1}$     & $|\varphi_{-}\rangle_{2}$   &
$|\varphi_{-}\rangle_{3}$  & $|\varphi_{-}\rangle_{4}$
&$|\phi^{-}\rangle_{P}|\psi^{-}\rangle_{S}$\\ 
$|\varphi_{-}\rangle_{1}$     & $|\varphi_{-}\rangle_{2}$   &
$|\varphi_{-}\rangle_{3}$  & $|\varphi_{-}\rangle_{4}$
&$|\psi^{-}\rangle_{P}|\psi^{-}\rangle_{S}$\\  \hline\hline
\end{tabular}\label{table2}
\end{table}

\begin{figure*}[th]
\centering
\includegraphics[width=18cm,angle=0]{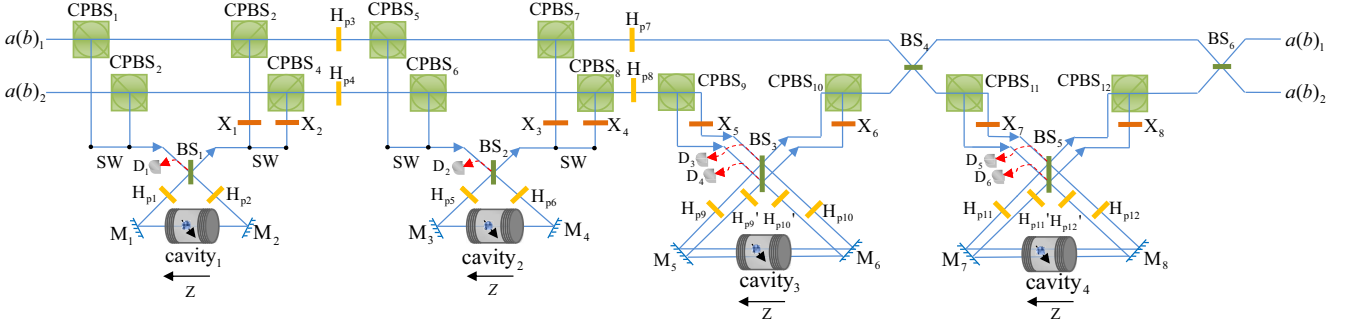}
\caption{Schematic diagram of the complete nondestructive
polarization-spatial HBSA.}  \label{fig6}
\end{figure*}

\section{Complete nondestructive polarization-spatial hyperentangled Bell-state analysis}

The protocol for the complete polarization-spatial HBSA proposed in
the former section is destructive. After the analysis of two-photon
polarization-spatial hyperentangled Bell states, the two photons are
detected at last and they cannot be used for other quantum
information processing tasks. In this section, we propose a complete
nondestructive polarization-spatial HBSA. The schematic diagram of
the complete nondestructive polarization-spatial HBSA is shown in
Fig.\;\ref{fig6}. The four QDs are all prepared in the states
$|\varphi_{+}\rangle_{1}=|\varphi_{+}\rangle_{2}=|\varphi_{+}\rangle_{3}
=|\varphi_{+}\rangle_{4}$ and the hyperentangled two-photon system
is in one of the $16$ hyperentangled Bell states.

One lets photon $A$ pass through the quantum circuit from the left
input port, followed by photon $B$. The interval time $\Delta t$,
which is less than the spin coherence time $\Gamma$, exists between
the two photons. After the two photons $A$ and $B$ passes through
quantum circuit, the evolution of the system composed of two photons
and four QDs is shown in Table\;\ref{table2}.

From Table\;\ref{table2}, one can obtain the complete and
deterministic analysis on hyperentangled Bell states of a two-photon
system $AB$ without detecting it. The measurement outcomes of
QD$_{1}$ and QD$_{3}$ show the parity information in the
polarization DOF and the spatial-mode DOF, respectively. In detail,
when QD$_{1}$ (QD$_{3})$ is in the state $|\varphi_{+}\rangle_{1}$
($|\varphi_{+}\rangle_{3}$), the two-photon system is in the state
$|\phi^{\pm}\rangle_{P(S)}$ in the polarization (spatial-mode) DOF.
Otherwise,  when QD$_{1}$  ($QD_{3})$ is in the state
$|\varphi_{-}\rangle_{1}$  ($|\varphi_{-}\rangle_{3}$), the
two-photon system is in the state $|\psi^{\pm}\rangle_{P(S)}$ in the
polarization (spatial-mode) DOF. The measurement outcomes of
QD$_{2}$ and QD$_{4}$ show the phase information in the polarization
DOF and the spatial-mode DOF, respectively. In detail, when QD$_{2}$
(QD$_{4}$) is in the state $|\varphi_{+}\rangle_{2}$
($|\varphi_{+}\rangle_{4}$), the two-photon system is in the state
$|\phi^{+}\rangle_{P(S)}$ or $|\psi^{+}\rangle_{P(S)}$ in the
polarization (spatial-mode) DOF. Otherwise, when QD$_{2}$ (QD$_{4}$)
is in the state $|\varphi_{-}\rangle_{2}$
($|\varphi_{-}\rangle_{4}$), the two-photon system is in the state
$|\phi^{-}\rangle_{P(S)}$ or $|\psi^{-}\rangle_{P(S)}$ in the
polarization (spatial-mode) DOF. That is, the schematic diagram
shown in Fig.\;\ref{fig6} can be used for the complete
nondestructive polarization-spatial HBSA in a  deterministic way.

\section{Discussion and summary}

Assisted by the optical transitions in a QD-cavity system, we
construct an error-detected block. With the error-detected block,
our schemes for the deterministic HBSG, complete HBSA, and complete
nondestructive HBSA of the polarization-spatial hyperentangled
two-photon system are proposed. In an ideal condition,
$|\Phi\rangle_{f1}=|L\rangle |i_{2}\rangle |\varphi_{-}\rangle$  or
$|\Phi\rangle_{f2}=|L\rangle |i_{2}\rangle |\varphi_{+}\rangle$ can
be obtained after the right-polarized photon $|R\rangle$ interacting
with the error-detected block and the fidelities and the
efficiencies of our schemes can be $100\%$. However, in a realistic
condition, the outcomes of the interaction between the
right-polarized photon $|R\rangle$ and the error-detected block,
which are described as Eqs.\;(\ref{eq9}) and (\ref{eq10}), are
affected by the coupling constant between $X^{-}$ and the cavity
$g$, the cavity decay rate $\kappa$, the leaky rate $\kappa_{s}$,
and the exciton dipole decay rate $\gamma$, which would affect the
fidelities and the efficiencies as well.

\begin{figure}[th]
\centering
\includegraphics[width=6.5cm,angle=0]{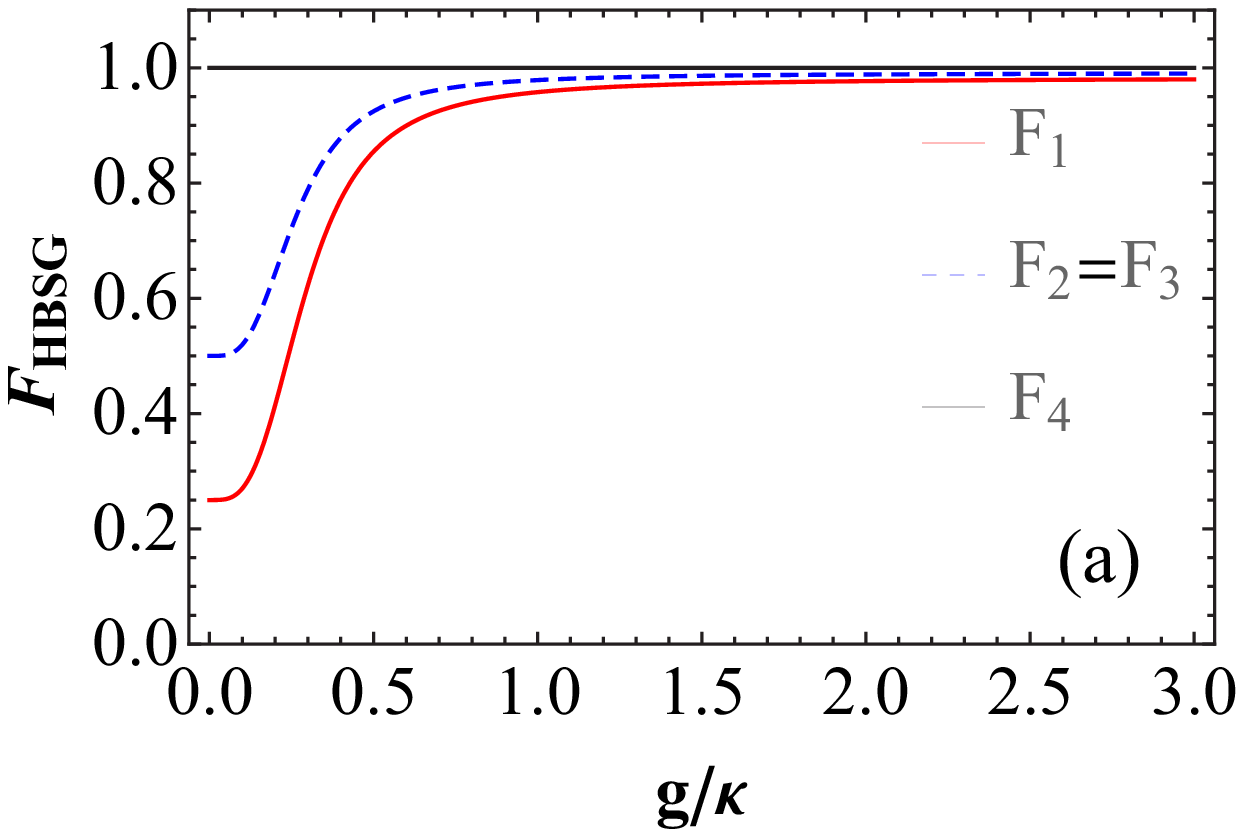} \hspace{20pt}
\includegraphics[width=6.5cm,angle=0]{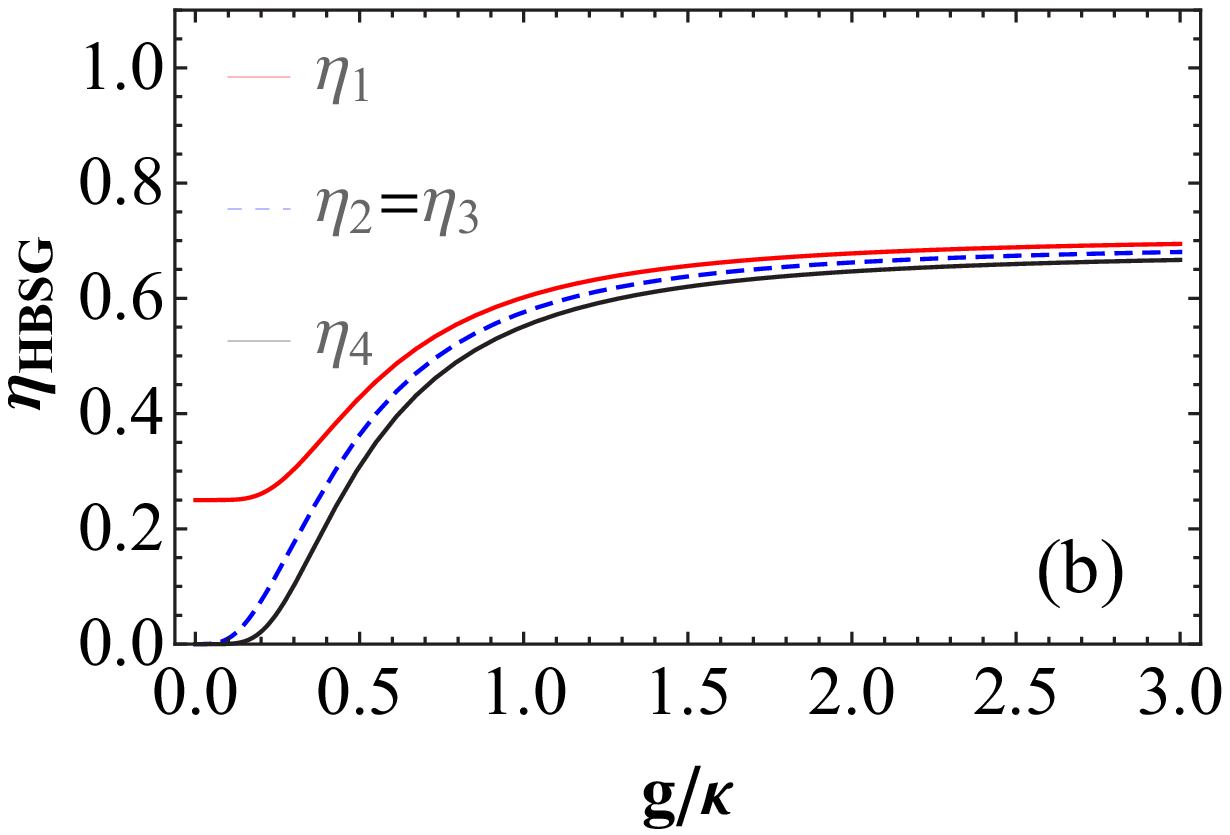}
\caption{ The performance of our deterministic HBSG scheme. The bule
dashed line, the red solid line, and the black solid line describe
the performance of our deterministic HBSG scheme for generating
hyperentangled Bell states
$|\phi^{+}\rangle_{P}|\phi^{+}\rangle_{S}$,
$|\phi^{-}\rangle_{P}|\psi^{+}\rangle_{S}$,
$|\psi^{+}\rangle_{P}|\phi^{-}\rangle_{S}$, and
$|\psi^{-}\rangle_{P}|\psi^{-}\rangle_{S}$, respectively. (a) The
fidelities of our deterministic HBSG scheme. (b) The efficiencies of
our deterministic HBSG. $\gamma=0.1\kappa$ and
$\kappa_{s}=0.2\kappa$, which are experimentally achievable, are
taken here.} \label{fig7}
\end{figure}

The fidelity of the process for deterministic HBSG, complete HBSA,
or complete nondestructive HBSA is defined as $F=|\langle \psi_{r} |
\psi _{i} \rangle|^{2}$. Here  $|\psi_{r}\rangle$ and
$|\psi_{i}\rangle$ are the final states in the realistic condition
and ideal condition, respectively.  The efficiency is defined as the
ratio of the number of the output photons to the input photons. The
fidelities of our scheme for deterministic HBSG are described as
\begin{eqnarray}    \label{eq17}
\begin{split}
F_{1}\;=\;&\frac{(T^{2}+1)^{4}}{4(T^{4}+1)^{2}},\\
F_{2}\;=\;&F_{3}\;=\;\frac{(1+T^{2})^{2}}{2(1+T^{4})},\\
 F_{4}\;=\;&1.
\end{split}
\end{eqnarray}
Here $F_{1}$, $F_{2}$, $F_{3}$ and $F_{4}$ are the fidelities of
generating the hyperentangled Bell states
$|\phi^{+}\rangle_{P}|\phi^{+}\rangle_{S}$,
$|\phi^{-}\rangle_{P}|\psi^{+}\rangle_{S}$,
$|\psi^{+}\rangle_{P}|\phi^{-}\rangle_{S}$, and
$|\psi^{-}\rangle_{P}|\psi^{-}\rangle_{S}$, respectively. The
fidelity of generating the hyperentangled Bell state
$|\psi^{-}\rangle_{P}|\psi^{-}\rangle_{S}$ is unity. The
efficiencies of generating these four hyperentangled states with our
scheme of deterministic HBSG are described as
\begin{eqnarray}    \label{eq18}
\begin{split}
\eta_{1}\;=\;&\frac{1}{4}(T^{4}+1),\\
\eta_{2}\;=\;&\eta_{3}\;=\;\frac{1}{2}(T^{2}+T^{6}),\\
\eta_{4}\;=\;&T^{4}.
\end{split}
\end{eqnarray}
The fidelities and the efficiencies of our deterministic HBSG scheme
vary with the parameter $g/\kappa$ in the conditions of
$\gamma=0.1\kappa$ and $\kappa_{s}=0.2\kappa$ are shown in
Figs.\;\ref{fig7}(a) and (b), respectively. From
Figs.\;\ref{fig7}(a) and (b), one can see that when $g/\kappa=0.5$,
which is a low-Q-factor of a cavity, the fidelities are
$F_{1}=85.45\%$ and $F_{2}=F_{3}=92.44\%$ and the efficiencies are
$\eta_{1}=42.79\%$, $\eta_{2}=\eta_{3}=36.32\%$, and
$\eta_{4}=30.83\%$ in the conditions $\gamma=0.1\kappa$ and
$\kappa_{s}=0.2\kappa$. For a strong coupling between the QDs and
the cavity, $g/\kappa>1$, the fidelities are $F_{1}>95.78\%$ and
$F_{2}=F_{3}>97.86\%$,  and the efficiencies are $\eta_{1}>60.17\%$,
$\eta_{2}=\eta_{3}>57.60\%$, and $\eta_{4}>55.13\%$ in the
conditions $\gamma=0.1\kappa$ and $\kappa_{s}=0.2\kappa$.

\begin{figure}[th]
\centering
\includegraphics[width=6.5cm,angle=0]{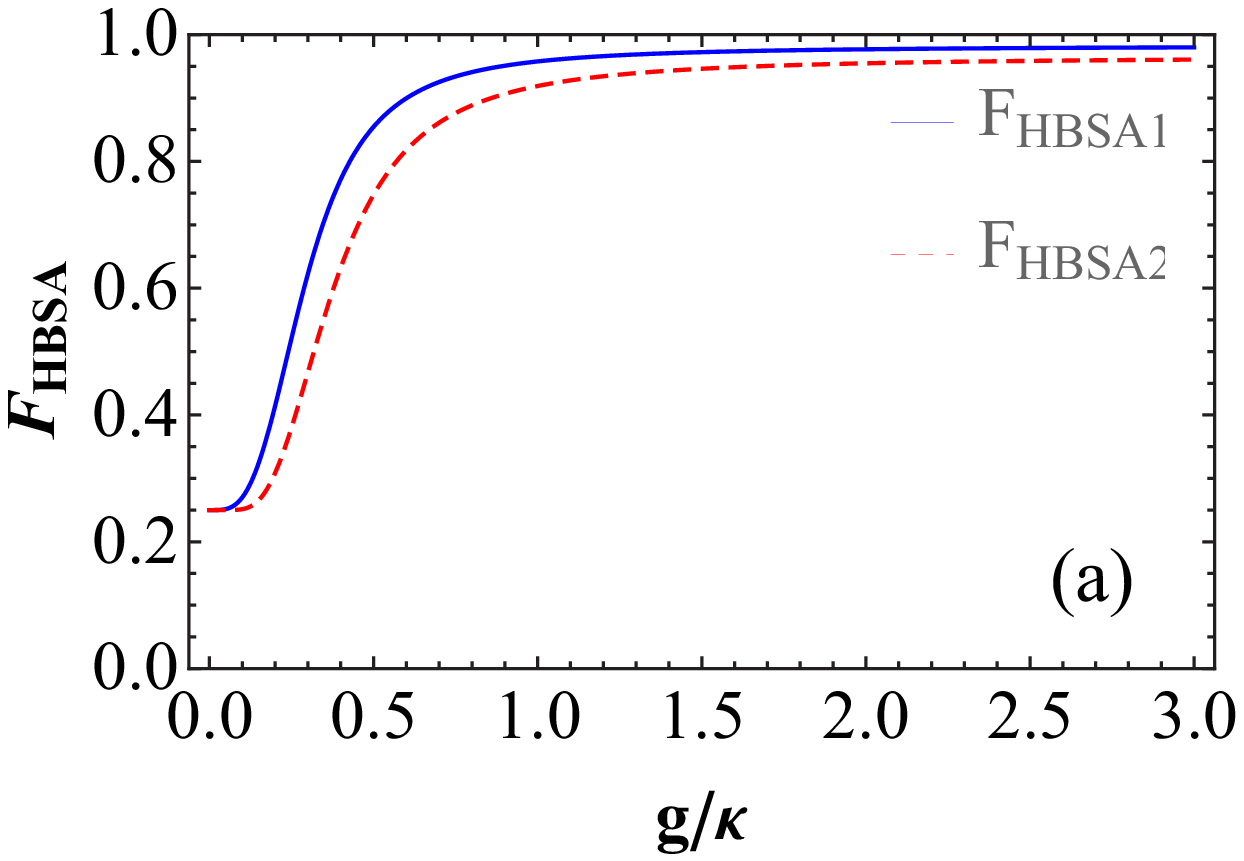} \hspace{20pt}
\includegraphics[width=6.5cm,angle=0]{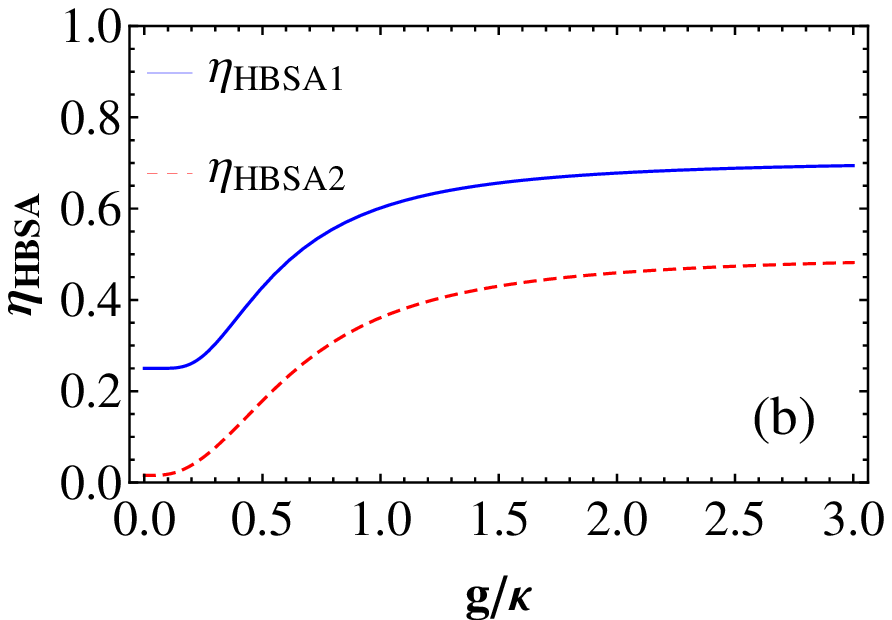}
\caption{ The performance of our complete HBSA scheme and complete
nondestructive HBSA scheme for the hyperentangled Bell state
$|\phi^{+}\rangle_{P}|\phi^{+}\rangle_{S}$. The blue solid line and
the red dashed line describe the performance of our complete HBSA
 and complete nondestructive HBSA schemes, respectively. (a) The
fidelities of our complete HBSA scheme and complete nondestructive
HBSA scheme. (b) The efficiencies of our complete HBSA scheme and
complete nondestructive HBSA scheme. $\gamma=0.1\kappa$ and
$\kappa_{s}=0.2\kappa$, which are experimentally achievable, are
taken here. } \label{fig8}
\end{figure}

The fidelity and the efficiency of our complete HBSA scheme for the
hyperentangled Bell state $|\phi^{+}\rangle_{P}|\phi^{+}\rangle_{S}$
are described as
\begin{eqnarray}    \label{eq19}
\begin{split}
&F_{HBSA1}=\frac{(T^{2}+1)^{4}}{(T^{2}+1)^{4}+2(1-T^{4})^{2}+(1-T^{2})^{4}},\\
&\eta_{HBSA1}=\frac{1}{16}[(T^{2}+1)^{4}+2(1-T^{4})^{2}+(1-T^{2})^{4}].
\end{split}
\end{eqnarray}
They vary with the parameter $g/\kappa$, shown as the blue solid
lines in Figs.\;\ref{fig8}(a) and (b), respectively. The fidelity
and the efficiency of our complete nondestructive HBSA scheme for
the hyperentangled Bell state
$|\phi^{+}\rangle_{P}|\phi^{+}\rangle_{S}$ are given by
\begin{eqnarray}    \label{eq20}
\begin{split}
&F_{HBSA2}=\frac{(1\!+\!T^{2})^{8}}{[(1\!+\!T^{2})^{4}\!+\!(1\!-\!T^{4})^{2}\!+\!4T^{2}(1\!-\!T^{2})^{2}]^{2}},\\
&\eta_{HBSA2}=\frac{1}{256}[(1\!+\!T^{2})^{4}\!+\!(1\!-\!T^{4})^{2}\!+\!4T^{2}(1\!-\!T^{2})^{2}]^{2}.\;\;\;\;
\end{split}
\end{eqnarray}
The red dashed lines in Figs.\;\ref{fig8}(a) and (b) describe
$F_{HBSG2}$ and $\eta_{HBSG2}$ varying with the parameter
$g/\kappa$, respectively. When the parameter $g/\kappa$ is larger
than $1$, the fidelities of our complete HBSA scheme and complete
nondestructive HBSA scheme will be higher than $F_{HBSA1}=95.78\%$
and $F_{HBSA2}=91.89\%$, respectively, for the hyperentangled Bell
state $|\phi^{+}\rangle_{P}|\phi^{+}\rangle_{S}$ in the conditions
$\gamma=0.1\kappa$ and $\kappa_{s}=0.2\kappa$. And the efficiencies
will be higher than $\eta_{HBSA1}=60.17\%$ and
$\eta_{HBSA2}=36.13\%$.

Besides the parameters $g$, $\kappa$, $\kappa_{s}$ and $\gamma$, the
fidelities would also be affected by the exciton dephasing,
including the optical dephasing and the spin dephasing of $X^{-}$.
Exciton dephasing reduces the fidelity by the amount of
$[1-exp(-\tau/\Gamma)]$, where $\tau$ and $\Gamma$ are the cavity
photon lifetime and the trion coherence time, respectively. The
optical dephasing reduces the fidelity less than $10\%$, because the
time scale of the excitons can reach hundreds of picoseconds
\cite{t1,t2,t3}, while the cavity photon lifetime is in the tens of
picoseconds range for a self-assembled In(Ga)As-based QD with a
cavity $Q$ factor of $10^{4}-10^{5}$ in the strong coupling regime.
The effect of the spin dephasing can be neglected because the spin
decoherence time is several orders of magnitude longer than the
cavity photon lifetime \cite{t4,t5,t6}.

In addition, the fidelity can be affected by the imperfect optical
selection induced by heavy-light hole mixing \cite{c62}. However,
this imperfect mixing can be reduced by engineering the shape and
the size of QDs or by choosing the types of QDs.

As the three protocols for deterministic HBSG, complete HBSA, and
complete nondestructive HBSA are constructed by our error-detected
block, the errors are detectable and the fidelities are improved.
The low efficiencies can be remedied by repeat until success. Once
our protocols succeeds, which means the detectors of the
error-detected block don't click, the high fidelities can be
obtained. This feature seems more important in the protocol for
deterministic HBSG, which means high-quality hyperentangled Bell
states can be generated.

In our schemes, the spin superposition state $|\varphi_{+}\rangle$
is prepared initially. It can be prepared from the spin eigenstates
by using nanosecond electron-spin-resonance pulses or picosecond
optical pulses. The preparation time for the spin superposition
state can be significantly shorter than $\Gamma$ because of the
ultrafast optical coherent control of electron spins in
semiconductor quantum wells and in semiconductor QDs. The
measurement of the spin superposition state in the basis
$\{|\varphi_{+}\rangle, |\varphi_{-}\rangle\}$ is required in each
scheme. By applying a Hadamard operation on the electron spin, the
spin superposition states $|\varphi_{+}\rangle$ and
$|\varphi_{-}\rangle$ can be transformed into the spin states
$|\uparrow\rangle$ and $|\downarrow\rangle$, which can be detected
by measuring the helicity of the transmitted or reflected photon.

In our proposal, the two cavity modes with right- and left- circular
polarizations, which couple to the two transitions
$|\uparrow\rangle\leftrightarrow|\uparrow\downarrow\Uparrow\rangle$
and
$|\downarrow\rangle\leftrightarrow|\downarrow\uparrow\Downarrow\rangle$,
respectively, are required. Many good experiments that provide a
cavity supporting both of two circularly-polarized modes with the
same frequency have been realized
\cite{modes1,modes4,modes6,modes7}. For example, Luxmoore \emph{et
al.} \cite{modes1} presented a technique for fine tuning of the
energy split between the two circularly-polarized modes to just 0.15
nm in 2012.

In summary, we have proposed some schemes for the deterministic
HBSG, the complete HBSA, and the complete nondestructive HBSA of the
polarization-spatial hyperentangled two-photon system assisted by
our error-detected block. The error-detected block is constructed by
the optical transitions in QD-cavity system. With the help of our
error-detected block, the errors can be detected, which is far
different from other previous schemes assisted by the interaction
between the photon and the QD-cavity system. Maybe this good feature
makes our schemes more useful in long-distance quantum communication
in the future.

\section*{Funding}
National Natural Science Foundation of China (11674033, 11474026,
11505007, 11547106);  Fundamental Research Funds for the Central
Universities (2015KJJCA01); Youth Scholars Program of Beijing Normal
University (2014NT28); Open Research Fund Program of the State Key
Laboratory of Low Dimensional Quantum Physics, Tsinghua University
(KF201502).

%
%

\end{document}